\documentclass{ifacconf}
\usepackage{graphicx}      
\usepackage{natbib}        
\usepackage{amsmath}
\usepackage{amsfonts}
\usepackage{gensymb}
\usepackage{glossaries-prefix}
\usepackage{subcaption}

\usepackage[textwidth=10mm]{todonotes}
\newacronym{pid}{PID}{proportional-integral-derivative}
\newacronym[prefixfirst={a\ },
            prefix={an\ }
            ]{ml}{ML}{machine learning}
\newacronym{drl}{DRL}{deep reinforcement learning}
\newacronym[prefixfirst={a\ },
            prefix={an\ }
            ]{nn}{NN}{neural network}
\newacronym[prefixfirst={a\ },
            prefix={an\ }
            ]{rl}{RL}{reinforcement learning}
\newacronym[prefixfirst={a\ },
            prefix={an\ }
            ]{mdp}{MDP}{Markov decision process}
\newacronym[prefixfirst={a\ },
            prefix={an\ }
            ]{mpc}{MPC}{model predictive control}
\newacronym{nmpc}{NMPC}{nonlinear model predictive control}
\newacronym{ahmpc}{AHMPC}{adaptive horizon model predictive control}
\newacronym{empc}{EMPC}{economic model predictive control}
\newacronym[prefixfirst={a\ },
            prefix={an\ }
            ]{mpcr}{MPC}{model predictive controller}
\newacronym[prefixfirst={a\ },
            prefix={an\ }
            ]{lqg}{LQG}{linear qudratic gaussian}
\newacronym{ocp}{OCP}{optimal control problem}
\newacronym{ou}{OU}{Ornstein–Uhlenbeck}
\newacronym{uav}{UAV}{unmanned aerial vehicle}
\newacronym{sac}{SAC}{soft actor critic}
\newacronym{msbe}{MSBE}{mean squared Bellman error}

\begin{document}
\begin{frontmatter}

\title{Reinforcement Learning of the Prediction Horizon in Model Predictive Control} 

\thanks[footnoteinfo]{This work was  financed  by grants from the Research Council of Norway (PhD  Scholarships  at  SINTEF grant no. 272402, and NTNU AMOS grant no. 223254).}

\author[First]{Eivind Bøhn} 
\author[Second]{Sebastien Gros} 
\author[First]{Signe Moe}
\author[Second,Third]{Tor Arne Johansen}

\address[First]{SINTEF Digital, Oslo, Norway (email: \{eivind.bohn, signe.moe\}@sintef.no}
\address[Second]{Department of Engineering Cybernetics, NTNU, Trondheim, Norway (email: sebastien.gros@ntnu.no)}
\address[Third]{Centre for Autonomous Marine Operations and Systems (e-mail: tor.arne.johansen@ntnu.no)}

\begin{abstract}
    \Gls{mpc} is a powerful trajectory optimization control technique capable of controlling complex nonlinear systems while respecting system constraints and ensuring safe operation. The \gls{mpc}'s capabilities come at the cost of a high online computational complexity, the requirement of an accurate model of the system dynamics, and the necessity of tuning its parameters to the specific control application. The main tunable parameter affecting the computational complexity is the prediction horizon length, controlling how far into the future the \gls{mpc} predicts the system response and thus evaluates the optimality of its computed trajectory. A longer horizon generally increases the control performance, but requires an increasingly powerful computing platform, excluding certain control applications. The performance sensitivity to the prediction horizon length varies over the state space, and this motivated the \gls{ahmpc}, which adapts the prediction horizon according to some criteria. In this paper we propose to learn the optimal prediction horizon as a function of the state using \gls{rl}. We show how the \gls{rl} learning problem can be formulated and test our method on two control tasks --- showing clear improvements over the fixed horizon \gls{mpc} scheme --- while requiring only minutes of learning.
\end{abstract}
\glsresetall


\end{frontmatter}

\section{Introduction}
\Gls{mpc} is a well studied and widely adopted control technique, particularly in the process control industry. Its popularity stems in large part from its ability to control complex systems while respecting system constraints, ensuring safe operation. It operates by solving an \gls{ocp} with the current state of the plant as the initial condition, and using a model of the plant to predict the plant response to the controlled variables. In this way it finds the sequence of control inputs that minimizes the objective function over the prediction horizon while remaining feasible in the sense of the trajectories remaining within the specified constraints. The first control input of the solution sequence is applied to the plant, and the \gls{mpc} then solves the \gls{ocp} again at the next sampling instance. The drawbacks of the \gls{mpc} framework is that the quality of the input sequence relies heavily on the accuracy of the model of the plant dynamics, the hyperparameters of the \gls{mpc} needs to be fine-tuned to the task at hand, and further that the computational complexity of solving the \gls{ocp} is fairly high, limiting the type of platforms and applications that can implement \gls{mpc}.

The prediction horizon length is a key parameter of the \gls{mpc} framework. In conjunction with the step size it controls how far into the future the controller evaluates the consequences of its actions. If chosen too short, the computed trajectories are myopic in nature and might lead to instability and poor approximations of the infinite horizon solution, while the computational complexity grows at best linearly with increasing prediction horizon. Moreover, different regions of the state space might have varying requirements on the horizon length for stability and to find nearly optimal trajectories. This observation motivated the \gls{ahmpc}. In \cite{michalska_robust_1993} the horizon is adapted so that a terminal constraint is satisfied and the system enters a known region of attraction of a second terminal controller. \cite{krener_adaptive_2018} proposes a heuristics-based approach, presenting one ideal but not implementable approach, and one practical method using iterative deepening search where stability criteria are checked on each iteration to determine the lowest stabilizing horizon. A more direct approach is presented in \cite{scokaert_min_1998} where the prediction horizon is included as a decision variable of the \gls{mpc} scheme. \cite{gardezi_machine_2018} proposes a learning based approach in which they construct a rich dataset of numerous combinations of states and \gls{mpc} computations with varying horizons, and then apply supervised learning on this dataset to develop an optimal horizon predictor.

\Gls{rl} \citep{sutton_reinforcement_2018} is a field of machine learning concerned with optimal sequential decision making. While \gls{rl} has proven to be the state-of-the-art approach for certain classes of problems such as game-playing \citep{schrittwieser2020mastering}, it has not seen many real world applications in control. This is in large part due to its data intensive nature, combined with its inability to handle constraints and therefore lack of guarantees for safe operation of the system, both in the learning stage and in production. However, \gls{rl} can be employed for control in a safe manner by using it to augment existing control techniques such as \gls{mpc} \citep{aswani2013provably,fisac2018general,zanon2020safe}, e.g. to learn the system dynamics \citep{nagabandi_neural_2018} or tune parameters \citep{mehndiratta2018automated}.

In this paper we propose to learn the optimal prediction horizon length of the \gls{mpc} scheme as a function of the state using \gls{rl}. To the best of our knowledge, this is the first work to employ \gls{rl} for \gls{ahmpc}. The contribution of this paper lies in exploring how the \gls{rl} problem of optimizing the \gls{mpc} prediction horizon can be formulated, and showcasing its effectiveness on two control problems. Further, we suggest to jointly learn the \gls{mpc} value function due to its synergistic relationship with the prediction horizon, enhancing the adaptive capabilities. While the \gls{ahmpc} approaches described earlier can be designed with favorable properties such as theoretical stability guarantees, they often assume access to privileged information such as terminal sets and control Lyapunov functions. Learned approaches on the other hand typically assume little is known, and as such are applicable to more problems.

The rest of the paper is organized as follows. Section \ref{sec:theory} presents the algorithms and theory employed in this paper. Section \ref{sec:method} presents the formulation of learning the optimal \gls{mpc} prediction horizon as \pgls{rl} problem, while Section \ref{sec:experiments} describes the experiments undertaken, the results of which are presented and discussed in Section \ref{sec:results}. Finally, Section \ref{sec:conclusion} concludes the paper with our thoughts about the proposed method and future prospects.

\section{Background}\label{sec:theory}
\subsection{Model Predictive Control}
\Gls{mpc} is a model-based control method where the control inputs are obtained by solving at every time step an open loop finite-horizon \gls{ocp} \eqref{eq:mpc}, using a model of the plant to predict the response to the control inputs from the current state of the plant. Solving the \gls{ocp} yields a control input sequence that minimizes the objective function over the optimization horizon. The first control input of this sequence is then applied to the plant, and the \gls{ocp} is solved again at the subsequent time step to get the next control input. In this paper we consider discrete-time state-feedback nonlinear constrained \gls{mpc} for which the \gls{mpc} receives exact measurements of the states at equidistant points in time. It reads as:
\begin{subequations}\label{eq:mpc}
    \begin{align}
        \min_{x, u} \quad &\sum_{k=0}^{N-1}\gamma^k \ell(x_k, u_k, \hat{p}_k) + \gamma^N m(x_N), \label{eq:mpc:obj}  \\ 
        \textrm{s.t. \quad} &x_0 = \bar{x} \label{eq:mpc:init} \\ 
        &x_{k+1} = f(x_k, u_k, \hat{p}_k), \quad \forall \enspace k \in 0, \dots, N - 1 \label{eq:mpc:dynamics} \\ 
        &H(x_k, u_k) \leq 0, \qquad \quad \ \ \forall \enspace k \in 0, \dots, N - 1 \label{eq:mpc:H}
    \end{align}
\end{subequations}
Where we here and in the rest of the paper use the notation $\min_*$ to indicate solving for the arguments that minimize the function. Here, $x_k$ is the plant state vector at optimization step $k$ and $\bar{x}$ is the plant state at the current time, $u_k$ is the vector of the control inputs, $\hat{p}_k$ are time-varying parameters whose values are projected over the optimization horizon, $f$ is the model dynamics, $H$ is the constraint vector and $N$ is the horizon. The state and control inputs are subject to constraints, which must hold over the whole optimization horizon for the \gls*{mpc} solution to be considered feasible. The \gls{mpc} objective function consists of the stage cost $\ell(x_k, u_k, \hat{p}_k)$, the terminal cost $m(x_N)$, and the discounting factor $\gamma \in (0, 1]$. The stage cost is problem specific, e.g. consisting of a tracking error and an input-change term $\Delta u_k^\top D \Delta u_k$ that discourages bang-bang control, where $\Delta u_k = u_k - u_{k-1}$. The stage cost only evaluates the trajectory locally up to a length of $N-1$ steps, the terminal cost $m(x_N)$ should ideally provide global information about the desirability of the considered terminal state, helping the \gls{mpc} avoid local minima. The more accurate the terminal cost is wrt. to the infinite horizon solution to problem \eqref{eq:mpc}, the shorter the horizon can be while still achieving good control performance \citep{zhong_value_2013,POLO}.

\subsection{Value Function Estimation}\label{sec:vf_estimation}
The ideal choice for the terminal cost $m(x_N)$ in the \gls{mpc} scheme would be the optimal value function $V^*$. A value function $V^\pi$ measures the expected total infinite horizon discounted cost accrued when following the control law $\pi$ \eqref{eq:valfn}, and the optimal value function is then the value of an optimal control law $\pi^*$ that chooses the optimal input at every point \eqref{eq:vstar}. Equation \eqref{eq:vstar} is written in the form of a Bellman equation where the value is decomposed into a one-step cost $\ell$ and the total value from the next state $x' = f(x, u, \hat{p})$. Computing $V^*$ exactly from \eqref{eq:vstar} is intractable for problems with continuous state and input spaces, and iterative approaches such as Q-learning requires an enormous amount of data for such problems. 

\begin{align}
    V^\pi(x, \hat{p}) &= \mathbb{E}\left[\sum_{t=0}^\infty \gamma^t \ell(x_t, \pi(x_t, \hat{p}), \hat{p}) \enspace | \enspace x_0 = x \right] \label{eq:valfn} \\
    V^*(x, \hat{p}) &= \min_u \ell(x, u, \hat{p}) + V^*(x', \hat{p}'), \enspace \forall x, \hat{p} \label{eq:vstar}
\end{align}

The \gls{mpc} scheme delivers local approximations to \eqref{eq:vstar}, and as such $V^\textrm{MPC}$ is a good surrogate for $V^*$ as the terminal cost $m(x_N)$. While computing $V^\textrm{MPC}$ exactly is not possible either --- due to requiring running the \gls{mpc} scheme with an infinite horizon --- it can be approximated with fitted value iteration from data gathered when running the \gls{mpc}.

\begin{subequations}\label{eq:val_fn_update}
    \begin{align}
        \theta^{\textrm{MPC}} &= \min_{\theta^{\textrm{MPC}}} \mathbb{E} \left[ \left(y(x, \hat{p}) - \hat{V}_{\theta^{\textrm{MPC}}} (x, \hat{p})\right)^2\right] \\
        y(x, \hat{p}) &=  \mathbb{E}\left[\ell(x, \pi^{\textrm{MPC}}(x, \hat{p}), \hat{p}) + \gamma \hat{V}_{\theta^\textrm{MPC}}(x', \hat{p}')\right]
    \end{align}
\end{subequations}

The value function approximator $\hat{V}_{\theta^{\textrm{MPC}}}$ is parameterized by the parameters $\theta^{\textrm{MPC}}$ that are updated according to \eqref{eq:val_fn_update} to minimize the \gls{msbe}. Moreover the \gls{mpc} scheme provides good approximations to the n-step Bellman equation, which when employed in the update rule \eqref{eq:val_fn_update} is known to accelerate convergence and promote stability of the value function learning process:

\begin{equation}
        V^*(x, \hat{p}) = \min_{u_0:u_{N-1}} \mathbb{E}\left[\sum_{t=0}^{N-1}\gamma^t\ell(x_t, u_t, \hat{p}_t) + \gamma^N V^*(x_N, \hat{p})\right] \label{eq:nstep}
\end{equation}

This is because a larger share of the value of the future trajectory is known exactly, and the contribution from the bootstrapping component $\gamma^N V(x_N, \hat{p})$ is reduced \citep{vanseijen2016effective,sutton_reinforcement_2018}.

\subsection{Reinforcement Learning}
The system to optimize using \gls{rl} is framed as a \gls{mdp} which is defined by a set of components $\left(\mathcal{S}, \mathcal{A}, \mathcal{T}, R, \gamma \right)$. $\mathcal{S}$ is the state space of the system, $\mathcal{A}$ is the action space, $\mathcal{T}$ is the discrete-time state transition function which describes the transformation of the states due to time and actions, i.e. $s' = \mathcal{T}(s, a)$, $R(s, a)$ is the cost function and $\gamma \in [0, 1)$ is the discount factor describing the relative value of immediate and future costs.

The aim of \gls{rl} methods is to discover optimal decision making for the problem as defined above, usually by constructing a policy $\pi_\theta$ --- i.e. a (possibly stochastic) function that maps states to actions, here parameterized by $\theta$ --- and/or a value function as in \eqref{eq:valfn}, where $\ell$ corresponds to $R$. The objective to be optimized is then:

\begin{equation}
    J(\theta) = \min_\theta \mathbb{E}\left[\sum_{t=0}^\infty \gamma^tR(s, \pi_\theta(s))\right], \forall s_0 \in \mathcal{S}_0
\end{equation}

that is, minimize the expected sum of costs acquired over the states visited by the policy in an infinite horizon. The expectation is taken over the initial state distribution $\mathcal{S}_0$, and the trajectory distribution generated by the policy and the state transition function.

\subsection{Soft Actor Critic}
\Gls{sac} \citep{haarnoja2018soft} is an actor-critic entropy-maximization \gls{rl} algorithm with a parameterized stochastic policy. Entropy is a measure of the randomness of a variable, and is in the case of a continuous action space defined as $\mathcal{H}(\pi_\theta(\cdot|s)) = \mathbb{E}_{a \sim \pi_\theta(a | s)}\left[- \log \pi_\theta(a|s)\right]$, i.e. the probability of taking a given action in state s given the policy $\pi_\theta$. In maximum entropy \gls{rl} the objective is regularized by the entropy of the policy, that is, the aim of the policy is to minimize the sum of expected costs while simultaneously maximizing the expected entropy. This in turn yields multi-modal behaviour and innate exploration of the environment, as well as improved robustness because the policy is explicitly trained to handle perturbations. For the sake of brevity, we will limit the discussion of the specifics of the \gls{sac} algorithm to the policy implementation, see \cite{haarnoja2018soft} for details on how the policy is optimized. \Gls{sac} learns a parameterized stochastic policy implemented as:

\begin{equation}
        \pi_\theta(s, \xi) = \tanh(\mu_\theta(s) + \sigma_\theta(s) \odot \xi) \label{eq:sac_policy}
\end{equation}

Here $\mu_\theta$ and $\sigma_\theta$ are the two outputs of the policy function approximator, representing the mean action and the covariance, respectively. $\xi \sim \mathcal{N}(0, 1)$ is independently drawn Gaussian noise, $\odot$ denotes element-wise matrix multiplication, and $\tanh$ is employed to squash the Gaussian's infinite support to the interval $[-1, 1]$. The policy can therefore control its entropy through the state-dependent noise covariance $\sigma_\theta$. When evaluating the policy we set $\sigma_\theta = 0$, such that the policy becomes deterministic, as this tends to give better performance.

\section{Method}\label{sec:method}
\subsection{Horizon Policy}\label{sec:method:horizon}
We learn a policy $\pi^N_\theta$ to output the prediction horizon $N$ of the \gls{mpc} scheme using \gls{sac}. The prediction horizon is a positive integer, that for convenience we choose to upper bound. As such we modify the output of the \gls{sac} policy by linearly scaling the output from the $\tanh$'s limits of -1 and 1, to 1 and $N_{\textrm{max}}$, and then round the output to the closest integer:

\begin{equation}
    a_t = \textrm{round}\left(\textrm{scale}\left(\pi^N_\theta(s_t), [-1, 1], [1, N_{\textrm{max}}]\right)\right) \label{eq:pi_horizon}
\end{equation}

The gradients of the \gls{rl} problem are not affected by these transformations as they are applied in the environment, while the gradients are calculated based on the unscaled and unrounded outputs from $\pi^N_\theta(s_t)$. This does however mean that the agent must ``learn" that similar outputs from the policy will be rounded to the same action in \eqref{eq:pi_horizon}, and thus lead to the same subsequent state and cost. We considered alternative ways of formulating the policy as a discrete distribution from which integer horizon lengths could be drawn directly, such as N-head \glspl{nn}, Poisson models, and negative binomial models, but settled on the described rounding approach due to its simplicity and favorable results.

The cost function of the horizon policy consists of a control performance cost $R_P$, i.e. the \gls{mpc} stage cost $\ell$, a constraint violation cost $R_C$, and a computation cost $R_N$ to encourage lower horizons when suitable: 

\begin{equation}
    R(s, a) = R_P(s') + \lambda_C (t_{\textrm{max}} - t) R_C(s') + \lambda_N R_N(a)
\end{equation}

where $\lambda_C, \enspace \lambda_N$ are weighting factors. $R_C(s)$ is a binary variable indicating whether a hard constraint of the problem was violated --- upon which the episode is ended --- and $t_{\textrm{max}} - t$ is the number of steps left in the episode such that the agent receives a penalty proportional to how early the episode is ended. We assume the computational complexity of the \gls{mpc} scheme grows linearly in the horizon length, i.e. $R_N(a) = a$, as a lower bound for the true complexity. This generally holds true for the interior point method we use if one assumes local convergence and an initial guess that is reasonable \citep{rao1998application}.

The \gls{rl} state space $\mathcal{S} = \left\{x, \hat{p}\right\}$ consists of the \gls{mpc} state space $x$ and the time-varying parameters $\hat{p}$, as these are necessary to ensure the Markov property.

\subsection{MPC Value Function}
The \gls{mpc}'s value function is trained jointly with the \gls{rl} horizon policy to minimize the \gls{msbe} as described in Section \ref{sec:vf_estimation}, using 32-step bootstrapping. We found that N-step learning provided sufficient stabilization such that other common techniques in value estimation such as target networks and multiple estimators were not needed \citep{fujimoto2018addressing}. We experimented with two types of approximators, \glspl{nn} and polynomial regression models, finding that they achieved similar prediction accuracy. We therefore use quadratic polynomial regression models due to their convexity, reducing the computational complexity of the \gls{mpc} scheme.

\subsection{Evaluation}
Since the environments are randomized we construct a test set consisting of 10 episodes for which all stochastic variables such as state initial conditions and references are drawn in advance and thereby fixed for all policies, ensuring a fair comparison. The learned horizon policy is compared against the standard \gls{mpc} scheme with a fixed horizon, to assess the contribution of the learning. Each fixed horizon \gls{mpc} also has its own value function estimated using a dataset of 15k time steps. 
\section{Experiments}\label{sec:experiments}
We illustrate our approach on two systems. We set $N_{\textrm{max}} = 50$, $\lambda_N = 3\cdot10^{-3}, 1\cdot10^{-3}$ and $\lambda_C = 10, 2$ for the inverted pendulum and collision avoidance systems, respectively. We use the hyperparameters suggested in the \gls{sac} paper \citep{haarnoja2018soft}, with the following exceptions: $\pi_\theta^N$ is a 2-layer fully connected \gls{nn} with 32 nodes in each layer, $\gamma = 0.97$ for both the \gls{mpc} scheme and the \gls{rl} algorithm, and a reward scaling of 0.6 for the inverted pendulum system and 0.3 for the collision avoidance system.

\subsection{Inverted Pendulum}
The first system we experiment on is the classic control problem of stabilizing an inverted pendulum mounted on a cart that is fixed on a track, so that the cart can only move back and forth in one dimension. The cart's position is constrained to the size of the track and the pendulum angle is constrained to be above perpendicular to the surface. The controller should also track a time-varying position reference. As the position of the cart and stabilization of the pendulum are intricately linked, respecting both of the constraints while tracking the position reference requires a fairly high optimization horizon. Each episode is terminated after a maximum of 100 time steps, or when a constraint is violated.

The state space consists of the states $x_k = \left[\eta, v, \beta, \omega \right]$, where $\eta$ and $v$ is position and velocity of the cart along the horizontal axis, while $\beta$ and $\omega$ is the angle to the upright position and the angular velocity of the pendulum. The system dynamics are described by the equations in \eqref{eq:cartpend}, where  $m = 0.2$ and $M = 0.8$ are the mass of the pendulum and total mass of cart and pendulum, and $l = 0.25$ is the length of the pendulum. For the \gls{mpc} model the dynamics are discretized with a step time of $\Delta_k = 0.04s$.

The stage cost $\ell(x_k, u_k, \hat{p}_k) = E_{\textrm{kinetic}} - E_{\textrm{potential}} + 10 \cdot (\eta_k - \eta_{k,r}) ^ 2 + 0.1 u_k ^2$ reflects the objective of stabilizing the pendulum in the up position, formulated through minimizing the negative potential energy of the system, while tracking the position reference $\eta_{k, r}$.

\begin{subequations}\label{eq:cartpend}
    \begin{align}
        \dot{\eta} &= v \label{eq:cartpend:pos} \\ 
        \dot{v} &= \frac{m g sin(\beta) cos(\beta) - \frac{4}{3}(u + m  l  \omega ^ 2 sin(\beta))}{m cos^2(\beta) - \frac{4}{3} M} \\
        \dot{\beta} &= \omega \\
        \dot{\omega} &= \frac{M  g  sin(\beta) - cos(\beta)  (u + m  l  \omega ^ 2  sin(\beta))}{\frac{4}{3} M  l - m  l  cos^2(\beta)} \label{eq:cartpend:omega} \\
        -5 &\leq u \leq 5, \enspace -1.5 \leq \eta \leq 1.5, \enspace -90\degree \leq \beta \leq 90 \degree 
    \end{align}
\end{subequations}

\subsection{Collision Avoidance}
The second system we consider is a reference tracking problem, in which a vehicle is controlled to follow a trajectory $\tau$ where obstacles are placed in the path that need to be avoided. The \gls{mpc} receives information about the reference trajectory as well as any obstacles in its vicinity, however the position of the obstacles grows more uncertain the longer the prediction horizon is. This means longer horizons considers increasingly uncertain information, and a short or medium horizon might be more suited in some situations. The episode is ended when reaching the end point of the trajectory, when colliding with an obstacle, or after a maximum of 150 time steps.

\begin{figure}
    \centering
    \includegraphics[width=0.4\textwidth]{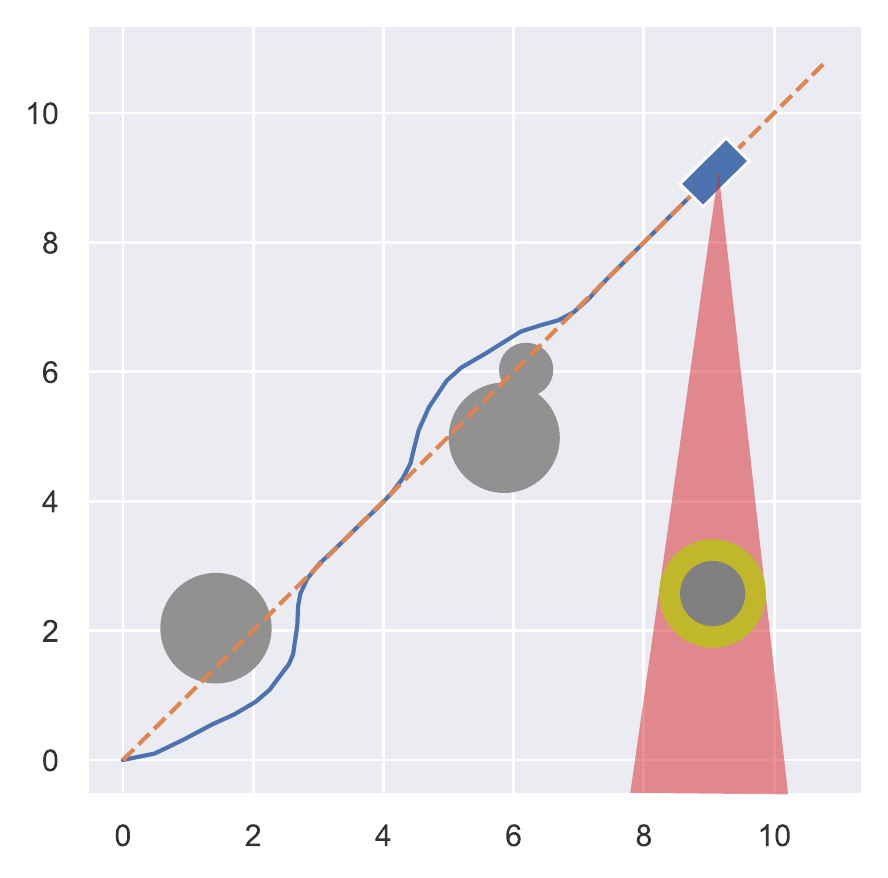}
    \caption{An episode in the collision avoidance environment. The blue rectangle represents the vehicle tracking the trajectory (the orange dashed line) from left to right, while avoiding the grey obstacles. The sensor beam's inaccuracy grows with distance, such that the projected position of the object in the beam is drawn from the yellow circle. The vehicles' size is enlarged for visual clarity.}
    \label{fig:ca_example}
\end{figure}

For the vehicle we employ a unicycle model \eqref{eq:unicycle}, where the \gls{mpc} provides a forward velocity $u_s$ as well as an angular velocity $u_\omega$ to turn the vehicle. The \gls{mpc} model is discretized with a step time of $\Delta_k = 0.1s$. The positions and sizes of the obstacles are randomly generated at the beginning of every episode, and their projected positions supplied to the \gls{mpc} are randomly drawn within a two-dimensional cone originating from the vehicle, such that the uncertainty grows the further away the object is from the vehicle. An episode is illustrated in Figure \ref{fig:ca_example}.

\begin{subequations} \label{eq:unicycle}
    \begin{align}
        \dot{p}_x &= u_s \cos(\beta) \label{eq:unicycle:px} \\
        \dot{p}_y &= u_s \sin(\beta) \\
        \dot{\beta} &= u_\omega \label{eq:unicycle:beta} \\
        0 &\leq u_s \leq 5, -4 \leq u_\omega \leq 4
    \end{align}
\end{subequations}

The stage cost is defined as $\ell(x_k, u_k, \hat{p}_k) = ||v_k - \tau_{k}||_2^2$ where $v_k$ and $\tau_k$ is the vehicle position and trajectory reference at time $k$. Further, soft constraints with slack variables are added around each obstacle with 150\% of the obstacles radius.

\section{Results}\label{sec:results}

\begin{figure*}[htpb]
    \centering
    \begin{subfigure}[b]{0.48\linewidth}
        \centering
        \includegraphics[scale=0.465]{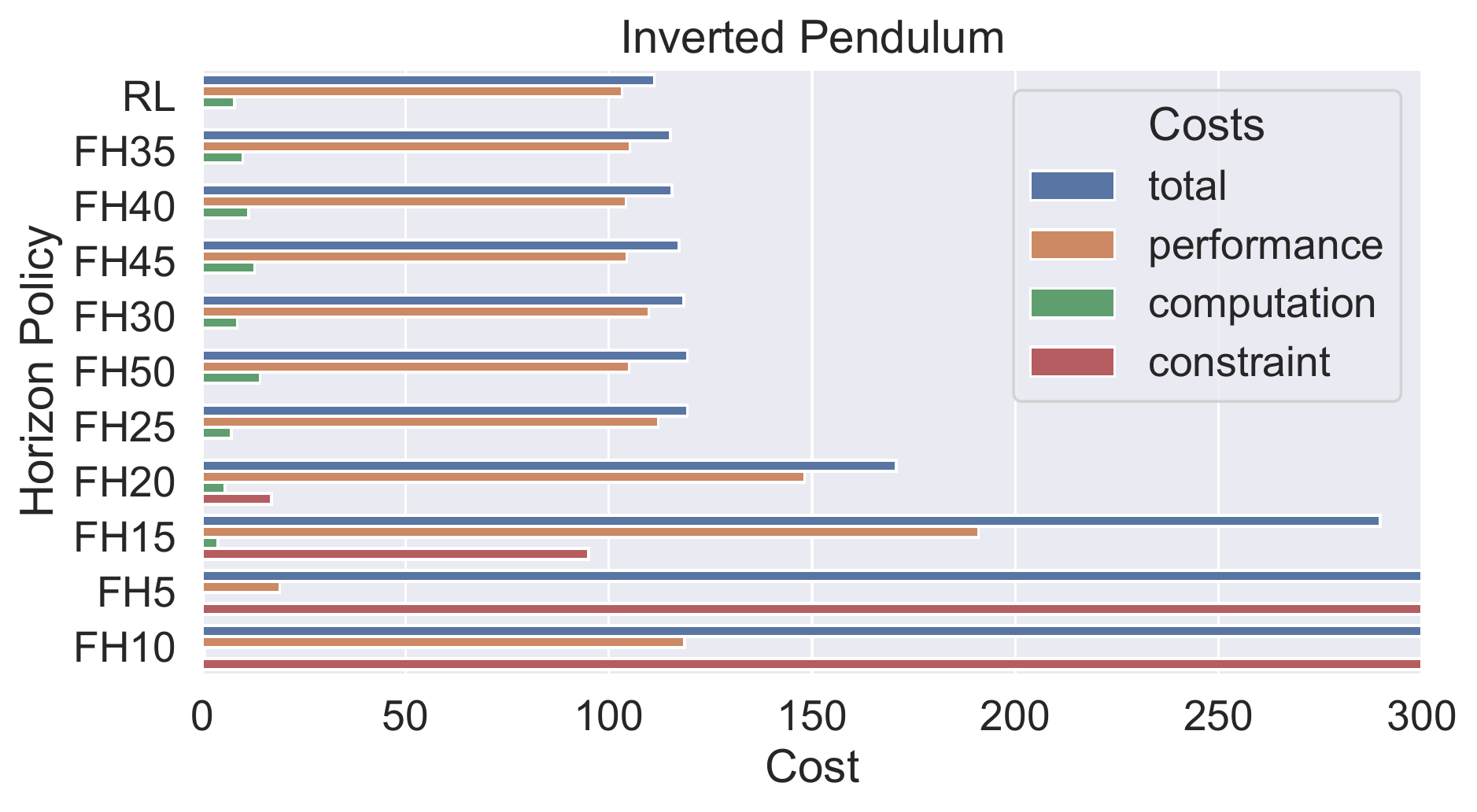}
    \end{subfigure}
    \begin{subfigure}[b]{0.48\linewidth}
        \centering
        \includegraphics[scale=0.465]{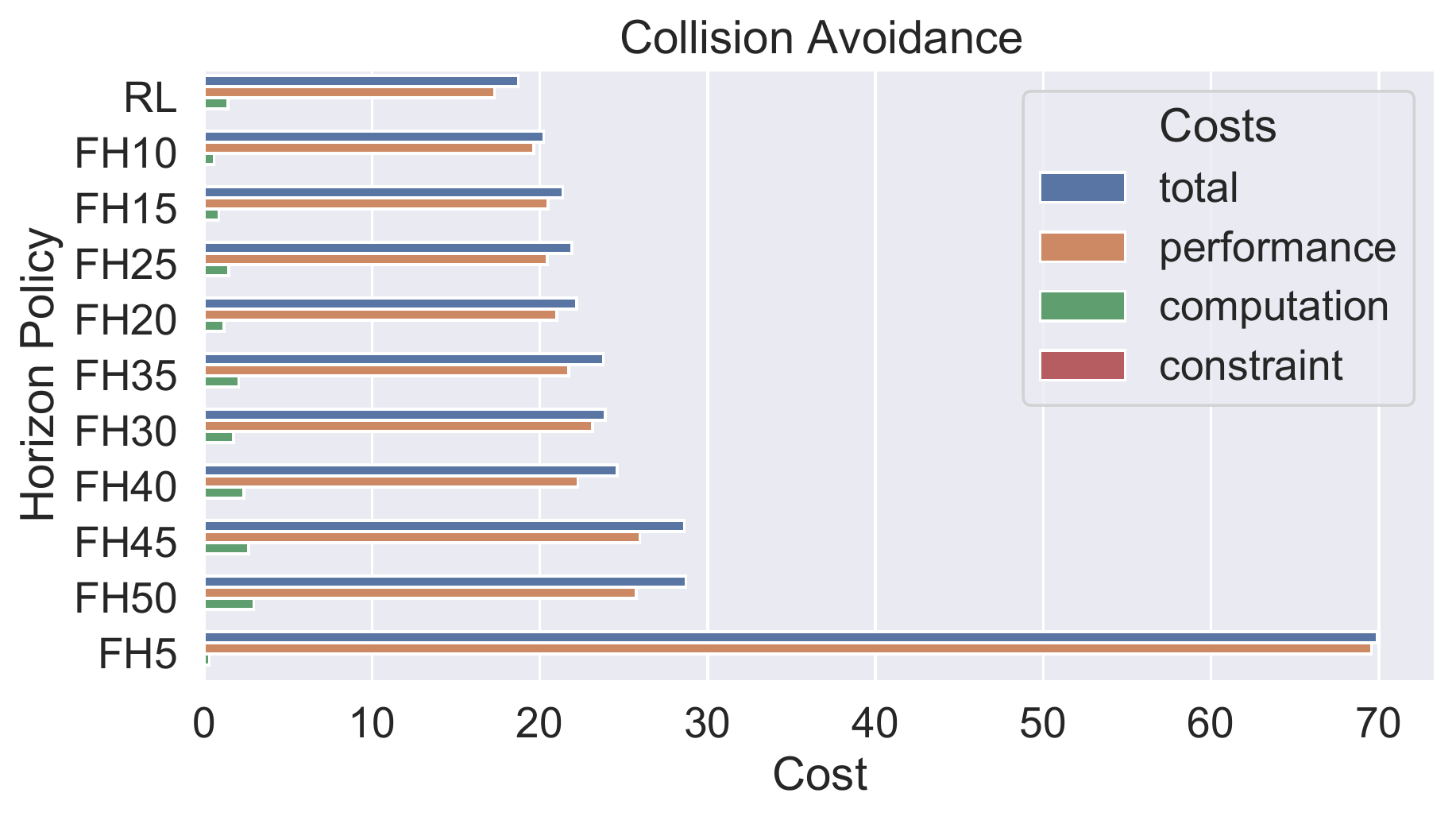}
    \end{subfigure}\label{fig:test-set:ca}
    \caption{Mean episode costs for the horizon policies on the test sets for the two systems using the value function as the terminal cost, where lower is better. The objectives are connected in that the policy does not accrue performance or computation cost after the episode is terminated from a constraint violation. The left figure is cut off due to the worst performing policies having a significantly higher cost.}
    \label{fig:test-set}
\end{figure*}

The standard \gls{mpc} scheme with various prediction horizons and the learned \gls{rl} \gls{ahmpc} are compared in Figure \ref{fig:test-set}. The \gls{rl} policy outperforms the standard \gls{mpc} scheme for all horizons lengths, improving on the second best achieving policy by about $4\%$ and $8\%$ for the inverted pendulum and collision avoidance systems, respectively. The improvement is more significant for the latter system as the performance objective varies more with the prediction horizons, and the \gls{rl} policy is able to identify when to use long and short horizons. For the inverted pendulum system, all horizons capable of respecting the constraints achieve similar performance costs, and as such the difference lies mainly in the computation term, although the \gls{rl} policy achieves the lowest performance cost here as well. \Gls{rl}'s ability to find improvement in a problem with such a noisy cost landscape and with such little potential improvement speaks to its strength. Moreover, the gains from reducing computation would be greater when using e.g. active set methods for the \gls{mpc} scheme which typically yields quadratic growth in computational complexity \citep{lau_comparison_2015}. 

\begin{figure}[htpb]
    \centering
    \begin{subfigure}[b]{0.48\linewidth}
        \centering
        \includegraphics[scale=0.4]{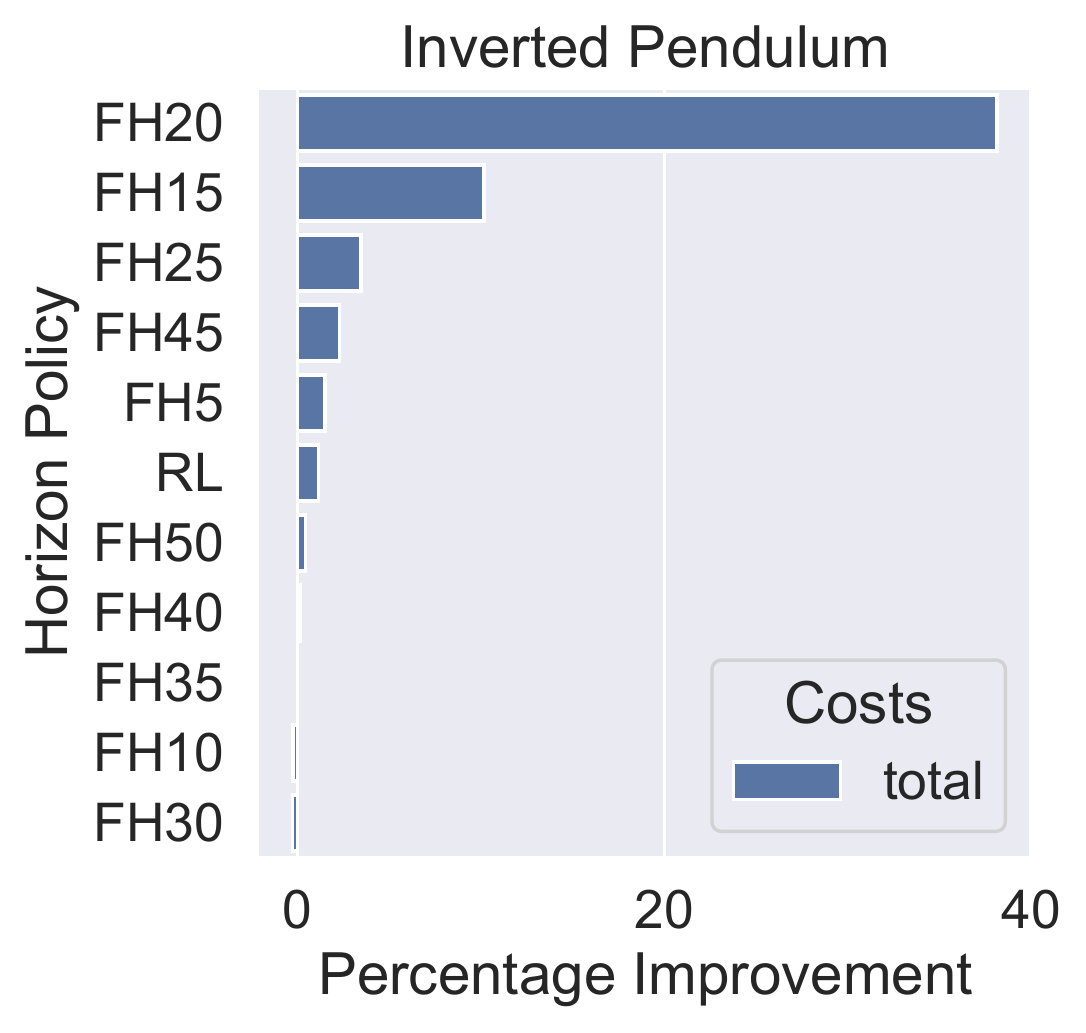}
    \end{subfigure}
    \begin{subfigure}[b]{0.5\linewidth}
        \centering
        \includegraphics[scale=0.4]{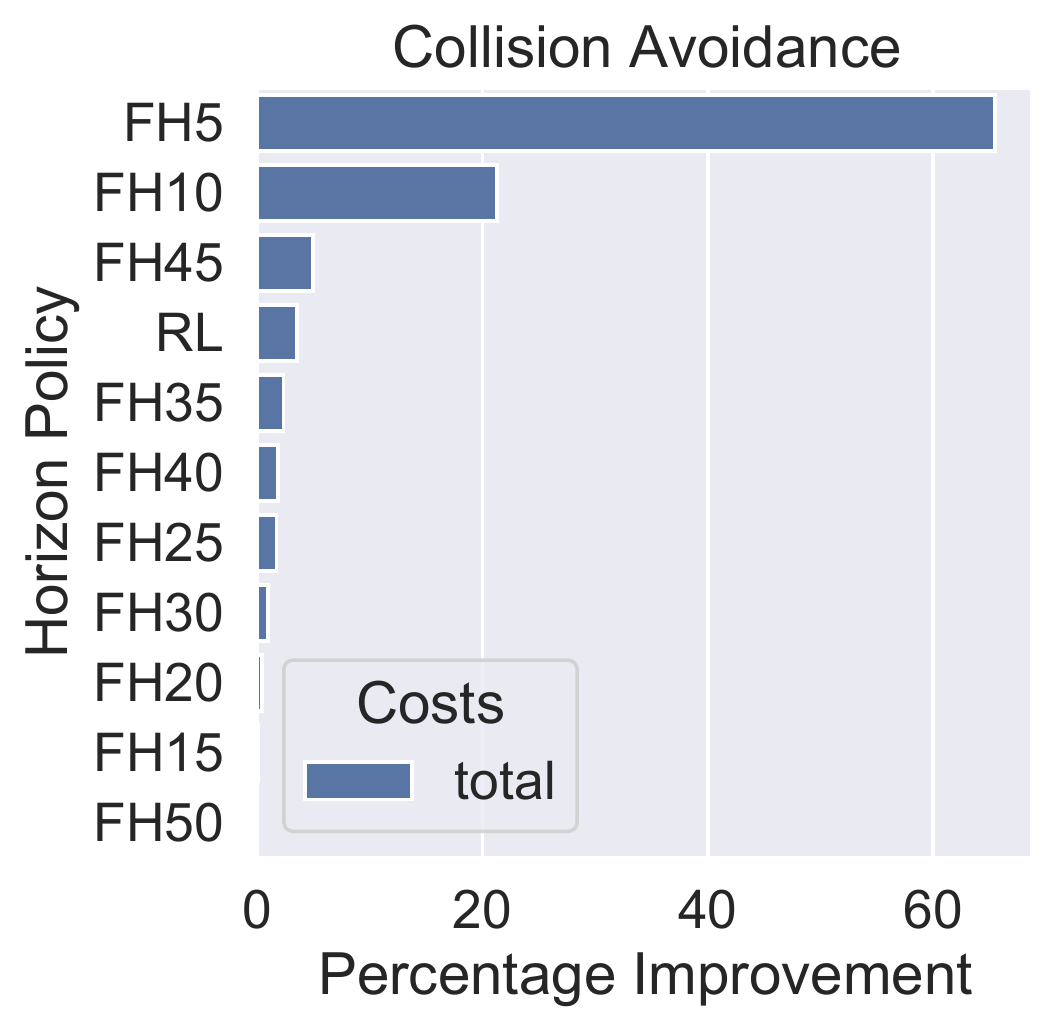}
    \end{subfigure}
    \caption{Total cost improvement over the test sets with the value function as the terminal cost in the \gls{mpc}.}
    \label{fig:vfdiff}
\end{figure}

\begin{figure}[htpb]
    \centering
    \begin{subfigure}[b]{0.48\linewidth}
        \centering
        \includegraphics[scale=0.35]{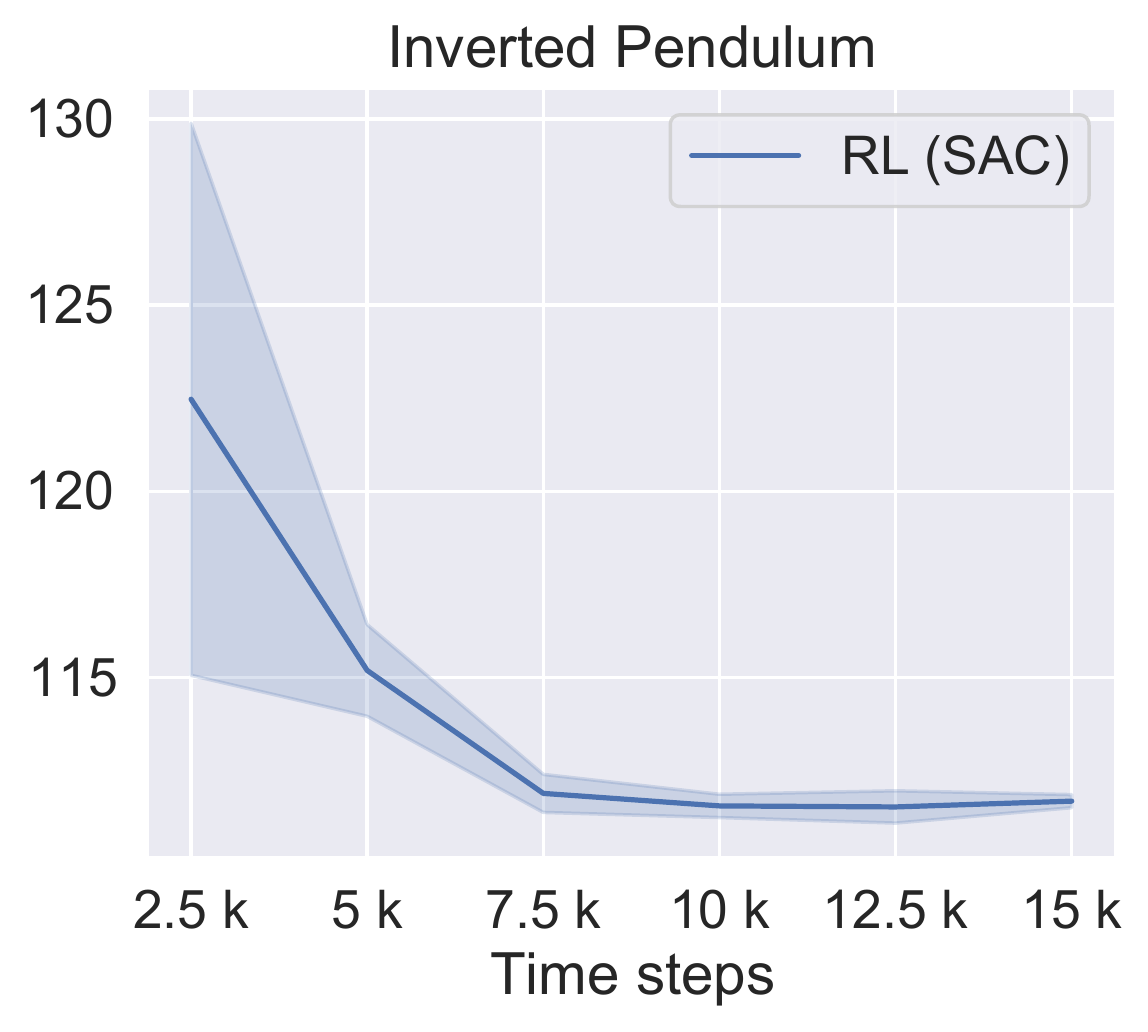}
    \end{subfigure}
    \begin{subfigure}[b]{0.5\linewidth}
        \centering
        \includegraphics[scale=0.35]{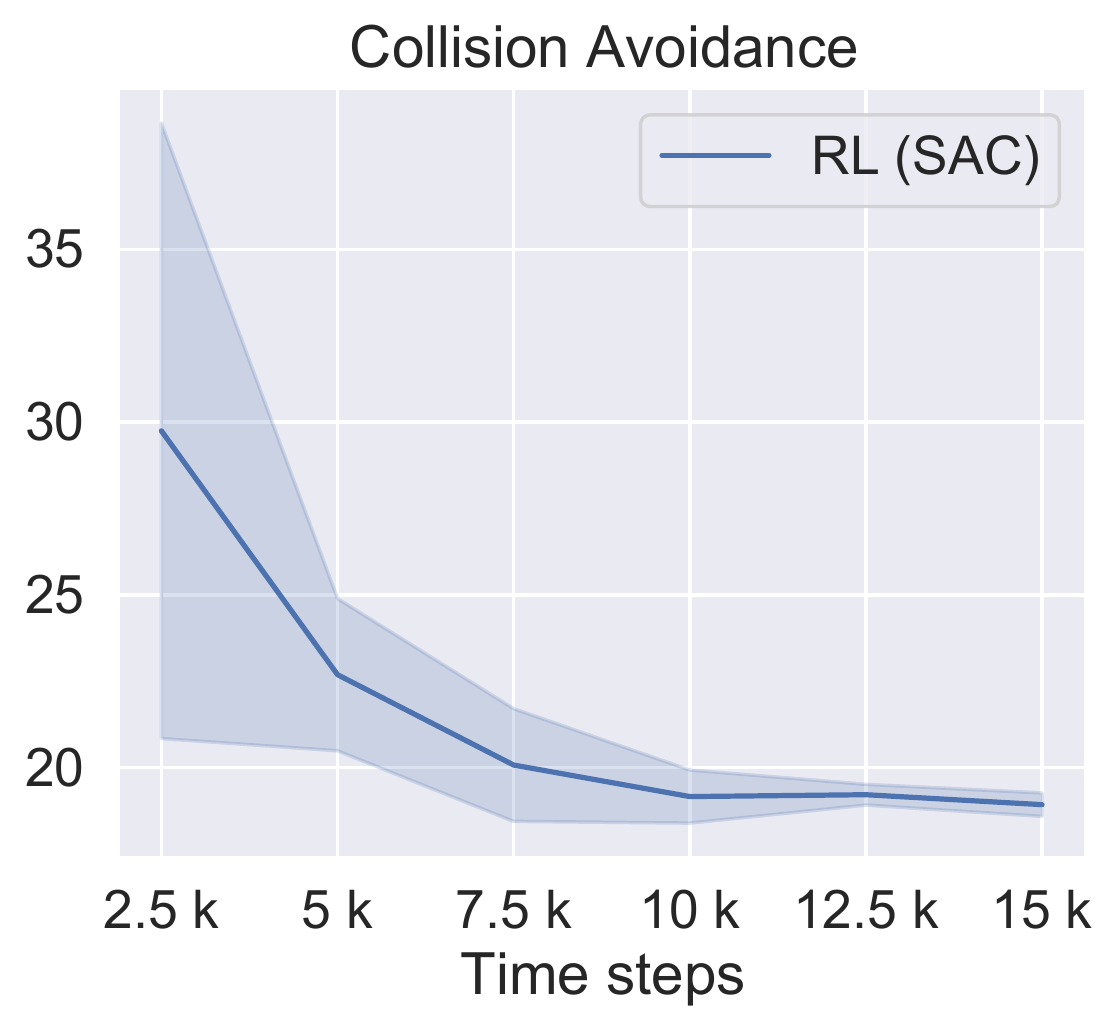}
    \end{subfigure}\label{fig:training:ca}
    \caption{Total cost on the test set for the \gls{rl} policy at different stages of the learning process. The solid line is the mean score while the shaded region is one standard deviation over three initialization seeds.}
    \label{fig:training}
\end{figure}

In the collision avoidance environment, the best performing fixed horizon is the short 10 step horizon. With the shortest 5 step horizon, the \gls{mpc} is unable to navigate around all the obstacles, preferring to stay still in front of large obstacles, although the addition of the value function mitigates this issue to some extent. Longer prediction horizons allows the \gls{mpc} to recognize that sometimes the longer way around the closest obstacle yields a shorter total path due to other obstacle locations, but its planned routes are more sensitive to the uncertainty in the projected locations. A robust \gls{mpc} scheme could alleviate this deficiency, however the \gls{rl} policy is also able to recognize this issue and leverage the strengths of both short and long horizons.

We found that implementing value function estimation in the \gls{mpc} scheme could significantly improve the performance when using horizons in a neighborhood of the horizon scale where performance changes abruptly, as illustrated in Figure \ref{fig:vfdiff}, which shows the percentage improvement for each policy when including $\hat{V}_{\theta^\textrm{MPC}}$ as the terminal cost. The \gls{rl} horizon policy does not benefit as much from the value function as we would expect, even being the best performing policy when removing the value function from it but not from the other policies. The benefit would probably be more significant in problems that are more temporally or spatially complex. In the collision avoidance problem the shortest horizons show the largest improvements, while for the inverted pendulum system the most improved horizons are the ones that lie close to the apparent minimum horizon required to successfully stabilize the pendulum and track the position reference. For both systems, the longer horizons benefit less from the addition of the value function. This is in part due to the fact that both these systems are heavily influenced by future information that is not available to the value function estimator, i.e. accurate information about distant obstacles and the future position reference for the cart. 

We note that the performance costs and the value function improvement is not monotonic wrt. the horizon length. This could partly be explained by the randomness in the data collection stage for the value function estimation.

Figure \ref{fig:training} shows the progression of the training process of the \gls{rl} horizon policy. It learns quickly, converging after around 15 thousand time steps for both systems, corresponding to about 10 and 25 minutes of data collection for the inverted pendulum and collision avoidance systems, respectively. Moreover, we find that the \gls{rl} horizon policy itself converges even faster and that the value function estimation is the slower, less data efficient component. From these results it seems evident that \gls{rl} is able to cope well with the rounding described in Section \ref{sec:method:horizon}.


\section{Conclusion}\label{sec:conclusion}
We have shown in this paper that \gls{rl} can be used to automatically tune and adapt the prediction horizon of the \gls{mpc} scheme on-line with only minutes of data collection, at least for simple systems. An important further work is to investigate how this affects the stability properties of the \gls{mpc} framework, and if any guarantees can be given.

\bibliography{references}

                                                   







\end{document}